\documentstyle[12pt,aasms4]{article}

\begin{document}

\lefthead{Carballo et al.}
\righthead{$K$-band imaging of B3-VLA quasars}

\newcommand{\kms}{{\rm km}\, {\rm s}^{-1}}
\newcommand{\col}{{\rm cm}^{-2}}
\newcommand{\lya}{Ly $\alpha$}
\newcommand{\lyb}{Ly $\beta$}
\newcommand{\erg}{{\rm erg}\, {\rm s}^{-1}\, {\rm cm}^{-2}}
\newcommand{\ergs}{{\rm erg}\, {\rm s}^{-1}}
\newcommand{\meneq}{\hbox{\raise 1 truemm \rlap{$<$}\lower 1 truemm 
\hbox{$\sim$}}}
\newcommand{\mayeq}{\hbox{\raise 1 truemm \rlap{$>$}\lower 1 truemm 
\hbox{$\sim$}}}

\title{$K$-band imaging of 52 B3-VLA quasars: Nucleus and host properties} 

\author{R. Carballo, S.F. S\'anchez}
\affil{Instituto de F\'\i sica de Cantabria (CSIC-Universidad de
Cantabria) and Departamento de F\'\i sica Moderna (Universidad de Cantabria). 
Facultad de Ciencias, 39005 Santander, Spain\\
Electronic mail: carballo@ifca.unican.es,
sanchez@ifca.unican.es}

\author {J.I. Gonz\'alez-Serrano}
\affil{Instituto de F\'\i sica de Cantabria (CSIC-Universidad de
Cantabria). Facultad de Ciencias, 39005 Santander, Spain\\
Electronic mail: gserrano@ifca.unican.es}

\author{ C.R. Benn}
\affil{Isaac Newton Group. Apartado 321, 38780 Santa Cruz de la Palma,
Spain\\
Electronic mail: crb@ing.iac.es}
\and
\author{M. Vigotti}
\affil{Istituto di Radioastronomia di Bologna, CNR, Via Gobetti 100, 40100
Bologna, Italy\\
Electronic mail: vigotti@astbo1.bo.cnr.it}

\date{1997 July 21}

\begin{abstract} 
    
We present $K$-band imaging and photometry of a sample of 52 radio loud
quasars (RQs) selected from the B3 survey with flux densities above 0.5 Jy
at 408 MHz. The optical completeness of the sample is 90\% and the quasars
cover the redshift range 0.4 -- 2.3. For $\sim 57$\% of the sources for
which the quality of the images allowed a detailed morphological study
(16/28) resolved extended emission was detected around the QSO, and its $K$
flux was measured. Interpreting this ``fuzz'' as starlight emission from
the host galaxy, its location on the $K-z$ plane at $z<1$ is consistent
with radio quasars being hosted by galaxies similar to radio galaxies (RGs)
or giant ellipticals (gEs). At higher redshifts the detected host galaxies
of RQs are more luminous than typical RGs and gEs, although some weak 
detections or upper limits are consistent with a similar fraction of RQs being 
hosted by galaxies with the expected luminosities for RGs or gEs.
 
The study of the $B-K$ colour distribution of the QSO nuclei, after
removing the contribution of $K$ emission from the host galaxy, confirm
that these sources are not reddened by large amounts of dust, with an
estimated extinction $A_{\rm V} < 1.0 $ mag at $z \simeq 1$.

We found a significant correlation between radio power and nuclear infrared
luminosity indicating a direct link between the radio synchrotron emission
and the nuclear emission in $K$. This correlation is more tight for the 
steep-spectrum sources (99.97\% significance). In addition, a trend is found
between radio power and infrared luminosity of the host galaxy (or mass),
in the sense that the most powerful quasars inhabit the most luminous
galaxies. The similarity of this tendency with that found for powerful
FR-II radio galaxies is consistent with the unification model for radio
sources.
\end{abstract}


\section{INTRODUCTION}

The B3-VLA Sample (Vigotti et al. 1989) is a catalogue of 1050 radio
sources selected at 408 MHz, consisting of five complete subsamples in
the flux density ranges 0.1-0.2 Jy, 0.2-0.4 Jy 0.4-0.8 Jy, 0.8-1.6 Jy
and $S_{408}>1.6$ Jy, and all mapped at the VLA in A and C
configurations at 1.46 GHz. From this catalogue Vigotti et al. (1997)
obtained the B3-VLA Quasar Sample, consisting of 125 sources with an
starlike counterpart in the POSS-I red plates with $R<20$ mag and
spectroscopically confirmed as quasars.  The sample covers the
redshift range $z=0.3-2.8$, with median redshift $z=1.16$, and radio
powers $P_{\rm 1.4 ~ GHz} \simeq 10^{27}-10^{28}$ 
WHz$^{-1}$ (the adopted cosmology in this work is 
$H_\circ=50$ km s$^{-1}$ Mpc$^{-1}$ and  $\Omega_\circ=1$).

The flux distribution of the Quasar Sample is as follows: 30 quasars with
0.1 Jy $<S_{408}<$ 0.4 Jy, 31 quasars with 0.4 Jy $<S_{408}<$ 0.8 Jy and 64
quasars with $S_{408}>0.8$ Jy.  The optical incompleteness of the sample,
i.e. the fraction of quasars fainter than the optical limit $R\simeq 20$
mag, depends on radio flux and is estimated to be around 6\% for
$S_{408}>0.8$ Jy, 10\% for $S_{408}>0.6$ Jy and 30\% for the flux bin
0.1-0.6 Jy (Vigotti et al. 1989, 1997; Benn et al. 1997, hereafter Paper
I).

Being selected at a low frequency most of the quasars have steep
spectrum.  The most notable radio sample selected at low frequencies
previous to the B3-VLA sample is the Third Cambridge Revised Radio
Sample (3CR, Spinrad et al. 1985; Laing et al. 1983), with optical
identifications and redshift measurements for all the sources.  The
3CR sample has a limiting flux of 9 Jy at 178 MHz and the median
redshift for the quasars is 0.7.  The B3-VLA Quasar Sample thus allows
the study of low frequency selected radio quasars (hereafter RQs)
reaching lower luminosities and higher redshifts than the 3CR sample.
A similar low-frequency selected sample is the Molonglo Reference
Catalogue/1 Jansky survey (MRC/1Jy, McCarthy et al. 1996; Kapahi et
al. 1996; Large et al. 1981), with a flux limit of 0.95 Jy at 408 MHz
and available optical identifications and redshifts for most of the
sources. We have started an observing programme aimed at the study of
the optical and near-infrared spectral energy distribution (hereafter
SED) of the B3-VLA quasars and the nature of their host galaxies.

One of the aims of the study of the optical to near-infrared SED of
the nucleus of these quasars was to determine the amount of reddening
due to dust absorption.  Being a radio-selected sample, it does not
have in principle any bias against obscured objects. The question
about the existence of obscured quasars and their number is crucial,
since a large fraction of obscured QSOs, such as the one claimed by
Webster et al. (1995), would imply profound revisions in
well-established properties of QSOs derived from optical surveys, such
as the quasar optical luminosity function and its evolution.  The
claim by Webster et al. (1995) of a large fraction of obscured quasars
was based in the red $B-K$ colours they found in a sample of
flat-spectrum PKS RQs, which they attribute to dust
reddening. However, as noted by Rieke, Lebofsky \& Wisniewski (1982)
and Serjeant \& Rawlings (1996), the red colours of flat-spectrum RQs
could be due to enhanced synchrotron emission, which extends to the
optical-infrared, due to relativistic beaming in a direction close to
the line of sight. The B3-VLA Quasar Sample, being predominantly a
steep-spectrum sample, is appropriate to determine the reddening using
the unenhanced (not beamed) core flux. Baker \& Hunstead (1995) and
Baker (1997) infer from the large Balmer decrements in Molonglo
lobe-dominated quasars extinctions up to $A_{\rm V} \simeq 3.7$, but
considerably less extinction is implied by the small reddening of the
continua; the maximum slope $\alpha_{\rm opt} \simeq - 2.2$ (the
spectral index $\alpha$ is defined as $S_\nu \propto \nu^{\alpha}$)
corresponding to $A_{\rm V} \simeq 1.4$ mag, assuming intrinsic
spectra $\alpha_{\rm opt} = -0.5$. The interpretation of the Balmer
decrements in the broad line regions of quasars is not
straightforward, as they depend on other effects apart from dust
reddening, such as radiative-transfer and collisional-excitation
effects (Osterbrock 1989, Baker et al. 1994). Furthermore it is not
clear that the same extinction applies to the broad emission lines and
the continuum.

Detailed studies of the host galaxies of active galactic nuclei (AGN)
are necessary to determine what kind of galaxies are able to ``feed''
an active nucleus, and to understand this phenomenon. The comparison
between the properties of the host galaxies of different classes of
AGN is crucial to explain the differences in nuclear activity. In
particular the comparison between the host properties of RQs and RGs
provides an effective test to the Unified Schemes for extragalactic
radio sources (see Antonucci 1993 for a review).

The general picture that has emerged from the study of RQ hosts in the
redshift ranges $z=0.2-1.0$ (e.g. Smith et al. 1986; Hutchings  1987;
V\'eron-Cetty \& Woltjer 1990; Dunlop et al. 1993, and their updated
work in Taylor et al. 1996; Disney et al. 1995; R\"onnback et
al. 1996) and $z=2-3$ (Lehnert et al. 1992, hereafter L92) is that
these galaxies have an elliptical morphological type (V\'eron-Cetty \&
Woltjer; Disney et al.; R\"onnback et al.; Taylor et al., hereafter T96)  
and large
sizes and $K$-band luminosities ($M_{\rm K} \simeq -26$; L92, T96), similar to those of giant ellipticals, which is also
the type of hosts harbouring RGs (Lilly et al. 1985; Lilly 1989;
Rigler et al. 1992; Best et al. 1997). The host galaxies of RQs
frequently exhibit morphological peculiarities in the form of tails
visible from the optical to the infrared, and a high incidence of
companions, suggesting interactions or merging processes in these
galaxies (Hutchings; Smith et al.; Disney et al.; Hutchings \&
Neff 1997). Tails seen in the infrared indicate in fact morphological
peculiarities in old stars, not related to nuclear radiation or star
formation. This characteristic is also shared with RGs, for which a
large fraction (50-75 \%) shows evidence of ongoing or past
interaction/merging processes (Heckman et al. 1986 for FRII powerful
RGs; Gonz\'alez-Serrano, Carballo \& P\'erez-Fournon, 1993 
for FRI low-luminosity RGs). Recently, optical structure coincident with
radio structure and interpreted as optical synchrotron was detected in
three 3C RQ hosts (Ridgway \& Stockton 1997). The flux contribution of the
synchrotron emission relative to the host galaxy is estimated to be around
10\% in the optical and lower than this value in the infrared.
 
The SED in the optical-infrared of several RQ hosts studied by L92 and
R\"onnback et al. (1996) is bluer than expected for elliptical
galaxies, although this was the type of galaxies implied by the
luminosity profiles and the $K-z$ relation. This result was also found
for RGs  (Lilly \& Longair 1982, 1984), which show a
rather wide range of optical-to-near-infrared colours. The wide range
of colours and the tight $K-z$ relation for RGs in the redshift range
$0.5<z<2.0$ were
explained by Lilly (1989, see also Rigler et al. 1992 and Lilly 1993)
assuming a two-component model comprising the cool and old red giant
stars evolved from the stellar population dominating the galaxy mass,
contributing more than 80\% of the $K$ emission, and a young
population of stars, i.e. a ``burst of star formation'' which explains
the excess emission in the optical and ultraviolet.
Detailed observations of the morphology of the optical continuum in RGs
(rest-frame ultraviolet) revealed that in most of the blue RGs at $z
\mayeq 0.8$ this emission is aligned with the radio axis, 
although the optical structure is not spatially coincident with the radio
structure  (McCarthy et al. 1987; Chambers, Miley \& van Breugel 1987).
The so-called ``alignment effect'' is one of the most intringuing properties
of RGs and the two most compelling explanations proposed have been 
scattered light from an anisotropically radiating nuclear source and 
star formation triggered by the radio jet (see McCarthy 1993 for a
review). Both mechanisms may operate, although the dominant process appears to
be scattering of nuclear light at $z\sim 1$ and induced star formations at
$z>3$ (Cimatti et al. 1997; Dey et al. 1997 and references in these
papers). The alignment is weaker for the infrared emission of RGs, which
is more symmetrical and nuclear concentrated, as expected for an old
stellar population (Rigler et al. 1992). 
These authors found that in typical 3C RGs at $z\sim 1$ the active
aligned component contributes about 10\% of the infrared light.
The infrared light profiles
of the $z \simeq 1$ RGs with only small blue components are in fact well
fitted by a de Vaucouleurs law (Rigler
\& Lilly 1994; Best et al. 1997), although presenting in some cases
excesses at large radii similar to those present in cDs (Best et
al. 1997). 

L92 and R\"onnback et al. (1996) suggest these same arguments of
recent star formation events or some contribution of scattered QSO
light at the higher frequencies to explain the blue colours of the RQ
hosts in their studies. We note that the work by L92 was biased
towards the selection of blue hosts, since L92 RQs were selected 
for the presence of extended UV-optical emission, and in fact all
contain extended Ly$\alpha$ emission (Heckman et al. 1991).
The sample used by R\"onnback et al. (1996) for their optical study of RQ 
hosts is not well defined. The B3-VLA Quasar Sample is appropriate to perform a
systematic study of RQ hosts using a well-defined sample, not biased
in principle towards the selection of blue objects. If the B3-VLA
quasars are hosted by large luminous galaxies similar to those found
by L92 and T96 they would be detectable in standard infrared images
obtained in 2-4 m telescopes.

In this work we present near-infrared imaging in the $K$ band of a
representative group of 52 quasars in the B3-VLA Quasar Sample.  We
selected the 47 quasars in this sample with $S_{408}>0.6$ Jy and right
ascensions in the range $7^{\rm h} - 14^{\rm h}$, so that they could
be observed in winter time, plus seven quasars with 0.5 Jy $<S_{408}<$
0.6 Jy and similar right ascensions. Two of the sources were excluded
from the study, since the identification of the quasar on the $K$
images was ambiguous. The optical completeness of this sample is
estimated to be around 90\% (Paper I and references therein).
The quasars cover the redshift range $z=0.4-2.3$ and the mean redshift
is $z=1.18$. This near-infrared database allowed us to address three
questions:

\begin{description}

\item{(i)} Loci of the 16 detected RQ hosts on the $K-z$ diagram   and
interpretation in terms of standard galaxy evolution models.

\item{(ii)} Analysis of the $B-K$ colours of quasar nuclei after fuzz
subtraction to impose limits on the relative amounts of obscuration in this
sample.  In Paper I we performed this study using the
total quasar $K$ magnitudes, uncorrected from the host galaxy emission.

\item{(iii)}  Infrared luminosities of host and quasar nuclei and their
relation with radio power.

\end{description}

The organization of the paper is as follows. The observations, standard
data reduction and $K$ photometry of the total quasar light is presented in
Sect. 2. In Sect. 3 we describe the technique used to separate the nucleus
and host galaxy component of the quasars and give the resulting magnitudes
of each component for the cases where both were detected. The issues (i),
(ii), and (iii) above are discussed in Sects. 4-6 and the summary and
conclusions are presented in Sect. 7.

\section {OBSERVATIONS AND DATA REDUCTION}

Near-infrared images of the 54 quasars were obtained on February the 5th
and 6th 1996 using the 256$\times$256 InSb infrared camera WHIRCAM at the
Nasmyth focus of the 4.2m WHT on La Palma (Spain). The $K_{s}$ filter was
used and the pixel scale was set to 0.27 arcsec pixel$^{-1}$, corresponding
to a field of view of $\sim$ $1'\times1'$.  The average seeing was $\sim 
1.9$ arcsec and $\sim 1.5$ arcsec during the first and second nights,
respectively. The $K_{\rm s}$ filter covers the wavelength range from 1.99 to
2.32 $\mu$m, with $\lambda_{\rm eff}$=2.16 $\mu$m. This filter is very similar
to Johnson $K$, and hereafter we will refer to it as $K$.

In order to avoid saturation due to atmospheric emission in the infrared
and obtain a high quality flat-field, the following observing procedure was
used. For each source, short unregistered exposures were taken and the
average image was registered. Five averaged images were obtained at five
different positions on the CCD, one near the centre and the rest at
symmetric directions at 9 arcsec offset relative to the first one. For
the targets we obtained 12 unregistered 10 s images per position and
for the photometric standards 5 frames of 4 s per position.  Care was
taken that the sources did not fall on a dead column near the left side of
the chip. A dark frame was obtained for every sequence of 5 registered
images in order to correct for the dark current.

The data were reduced using standard tasks in the IRAF package. First,
from each of the five registered images we subtracted the
corresponding dark frame. A flat-field frame was obtained for each
object using the average of the images obtained at the five different
positions. This average was made clipping out the highest intensity in
each pixel in order to remove the contribution from sources and
cosmic-ray events. Each of the five images in the
sequence was then corrected using this flat-field, and the final image
was obtained as the average of the 5 flat-fielded images shifted to a
common central position. The total exposure time of the final target
frames was of 600 s.

All the images were obtained in good photometric conditions. Flux
calibration for each night was carried out using UKIRT
standard stars from the WHIRCAM User's Guide.  Photometric calibration was 
better than 0.1 mag. The surface brightness level reached in the
images is $\mu_K \simeq 22.6$ mag arcsec$^{-2}$ and the 3$\sigma$
limiting magnitude for point sources is $K \simeq 19$ mag.

All the quasars were detected on the images. For two objects (0937+391
and 1256+392) two possible counterparts were detected on the $K$
images, both consistent with the optical/radio position, and it was
impossible to tell which one was the right counterpart.

$K$ magnitudes of the detected quasars were measured on the images
using circular apertures centred at the emission peak. 
Aperture diameters ranged from
5.5 to 12 arcsec depending on the seeing, collecting $\sim 100$\% of the
light.  Typical $K$ magnitude errors are around $\pm 0.1$
magnitudes.  The $K$ magnitudes are listed in Table 1, along with some
optical and radio information about the quasars.  The radio data, from
Vigotti {\it et al.} (1989), include the total flux density at 408
MHz and 1460 MHz, and the spectral index
$\alpha_{408}^{1460}$. The redshifts were taken from Vigotti et al. 
(1997). The $B$ and $R$ optical magnitudes in Table 1 
were taken from the catalogue of objects on the POSS-I blue and red
plates generated by the Automatic Plate-measuring Machine (APM) in
Cambridge (Irwin 1992), and were corrected from Galactic extinction. 
The distribution of $B$ and $K$ magnitudes versus redshift of the
B3-VLA quasars is shown in Figure 1.  Crosses correspond to
flat-spectrum sources, defined by $\alpha_{408}^{1460} > -0.5$. 

\section {ANALYSIS OF THE DATA}

The main problem for the detection and photometry of quasar host
galaxies is that the active nucleus dominates the quasar light in a
large range of wavelengths, making difficult the substraction of the QSO
from the image. As noted by Dunlop et al. (1993), 
at wavelengths around 1 $\mu$m, the
SED of a normal galaxy has a maximum whereas that the SED of an active
nucleus has a minimum (Sanders et al. 1989; Elvis et al. 1994).
Therefore this is the best wavelength range to detect galaxy hosts,
where the ratio of the nuclear to host galaxy emission is minimum. See
Fig. 1 of McLeod \& Rieke (1995) for the comparison between the SED
of a typical galaxy and that of an active nucleus. At the mean redshift of the
B3-VLA quasars of our study, $z=1.18$, this rest-frame wavelength
corresponds precisely to the $K$ band. Another advantage in the use of
the near infrared, compared to the optical, is that the galaxy
infrared emission arises from the old stellar population, related to
the galaxy mass. In fact some quasars show optical extended emission
which is originated by transient phenomena in the gas, through its
interaction with the active nucleus (Heckman et al. 1991).

In this work the quasar images were analysed using the following
procedure.  First surface brightness profiles were obtained from the
deconvolved images. Then these profiles were fitted with a
two-component model describing the nuclear source and the galaxy.  In
fact the analysis assumes, as working hypothesis, that any detected
extended emission component at $K$ is stellar light from the host
galaxy, and hence we used galaxy models to describe the extended
emission. The contribution of
extended emission in the $K$ band due to an aligned component similar to
that in RGs or beamed optical synchrotron emission is expected to be lower
than 10\% (Rigler et al. 1992; Ridgway \& Stockton 1997). 
A fraction of nebulosities around quasars show strong UV/optical
emission lines (Boroson \& Oke 1984; Boroson, Persson \& Oke 1985;
Stockton \& Mackenty 1987) which may reach equivalent widths of $\sim
500$\AA\, but typically $\sim 100$\AA. In the $K$ filter we may have
contribution of H$\alpha$ emission from the fuzz 
for the three objects in the redshift range
$2 < z < 2.3$, but from these equivalent widths we estimate a maximum 
contribution to the $K$ flux of the fuzz around 15\%, 
which is the fraction for the
nebulosities with the strongest emission lines.

Due to some technical failure all the second night images are slightly
out of focus, preventing the detailed analysis needed for the
detection of the host galaxies. For this reason we restricted the
search for host galaxies to the first night images (38 quasars). This 
problem with the second night images does not affect the quasar
aperture $K$ magnitudes presented in Table 1.

\subsection {Point spread function (PSF) and image deconvolution}

Atmospheric and instrumental point spread function has important effects
not only on the spatial resolution but  also on the observed surface
brightness profiles of galaxies (Capaccioli \& de Vaucouleurs 1983; Bailey
\& Sparks 1983) affecting mainly the central regions. The effect is
specially strong 
 when the galaxy has a bright point-like nucleus. As we are interested
in measuring host galaxy fluxes, deconvolution allows us, not only a better
separation of the nuclear component, but also a more accurate determination
of  galaxy and nuclear fluxes.

Due to the small field of view of the images, $1' \times 1'$, only a few
objects have stars in the frame, and in most cases, these are faint. We
selected the four brightest field stars, distributed evenly through the
first night, to built a normalized PSF. The FWHM of the PSF is 1.9 arcsec
and it represents the average seeing of the night, which had variations of
$\pm$0.3 arcsec (1 pixel). We used the Lucy-Richardson algorithm (Lucy
1974; Richardson 1972) and this PSF to deconvolve the images.  Because of
seeing variations and scarcity of good field stars this is not an ideal PSF
for the whole night data.  With a perfect PSF determination and infinite
signal-to-noise data, the deconvolution process would produce a narrow
($\sim$ 1 pixel width) point source surrounded by some diffuse emission if
this was present.  However, it has been shown that there is a resolution
limit due to photon noise which prevents the deconvolved PSF of being of
zero FWHM (Lucy 1992).  In our data, with moderate signal-to-noise and
using an average FWHM, the mean FWHM of the stars after deconvolution is
$1.3\pm0.3$ arcsec, with the dispersion reflecting the seeing variations
and the intrinsic dispersion from the deconvolution process itself. It is
important to stress that although there is not a large gain in spatial
resolution, the deconvolution technique allows a better determination of
the nuclear and extended fluxes.

\subsection {Surface brightness profile fitting}

Surface brightness profiles have been obtained for the deconvolved quasar
images using the techniques discussed in Jedrzejewski (1987). In brief,
given an initial value for the centroid of the object, the light
distribution is sampled along a first-guess elliptical isophote.  This
produces a one-dimensional intensity distribution as a function of the
eccentric anomaly. This distribution is then analyzed as a harmonic
expansion in a Fourier series and finally, the parameters of the isophote
(position angle, ellipticity) are estimated by a least-squares fit. This
procedure is repeated by increasing the semimajor axis of the ellipses
until the sky background level is reached. The final product consists of
semimajor-axis profiles of intensity, ellipticity and position angle. The
resulting brightness profiles were fitted using an interactive least-square
method. Four different models were used: (i) gaussian function; (ii)
gaussian + King profile; (iii) gaussian + $r^{1/4}$ law; (iv) gaussian +
exponential profile.  The results of the analysis for the stars in the
images showed that a gaussian function is the best representation of the
PSF.

The original images of some of the quasars (10) 
were very noisy or they had
(or could have) problems of bad tracking or a slight out-of-focus and
we did not try to deconvolve them, although two of them
(0922+425 and 1258+404) showed structure that could be real.
The structure could have arisen because
the host is part of a galaxy system, has multiple nuclei, or
gravitational lensing produces several images of the quasar. We did
not obtained the surface brightness profile of these 10 sources.

For the fitted surface brightness profiles each best fit provided us
with two or four parameters: two for the central point source and two
for the extended component if this was used. For the cases where we
fitted a gaussian plus an extended component the QSO magnitudes
($K_{\rm QSO}$) were obtained directly from the model and the host
magnitudes ($K_{\rm gal}$) were obtained from the difference between
the total aperture flux ($K$ magnitudes in Table 1) and the QSO flux
from the model.

In order to set the reliability of the method we have applied the whole
procedure (deconvolution, determination of the surface brightness profile
and profile fitting) to 12 stars on the frames.  This test is useful to
quantify the effect of using an average PSF. In addition, the deconvolution
algorithm could enhance the wings of bright sources, since the noise in the
outer parts is high, and spurious detections of extended emission around
point sources could arise because of this effect. Although for some of the
stars a two-component model gave a good fit, the extended component never
contributed more than 12\% to the total flux. This fraction corresponds to
$K_{\rm gal} \sim 17.9$ mag for the average magnitude $K=15.6$ of the
quasars in the sample.  This magnitude is approximately similar to the
3$\sigma$ limiting magnitude of an extended source of size of $\sim 10$
arcsec, which is the typical size over which the fits were obtained. We
used this magnitude, $K=17.9$, as an absolute upper limit for the
reliability of the detection of an extended component. For the quasars
brighter than 15.6 the limit of 12\% fixes a lower limiting magnitude for
galaxy detection, corresponding to $K_{\rm gal} = 16.3$ for the brightest
quasars with $K=14$. A fractional contribution from the galaxy of 12\%
gives only a correction of 0.14 mag to the quasar $K$ magnitude.

For twelve quasars a pure gaussian model produced a good fit giving
small rms (less than 0.1 mag and typically $10^{-3}$ mag), and those
were classified as point sources. These quasars are 0701+392,
0827+378, 0923+392, 1105+392, 1128+385, 1144+402, 1148+387, 1203+384,
1206+439B, 1312+393, 1339+472 and 1343+386.  For the remaining 16 objects
we found  much larger residuals, clearly deviating from point sources, and
their profiles were therefore fitted  
using models (ii), (iii) and (iv) explained
above. Figure 2 shows an example of a fit with a $r^{1/4}$ galaxy
model. For all of these cases the residuals for two or the three
two-component models were smaller than the residuals for the gaussian
fit, and the rms values for these fits were roughly similar. For this
reason we could not determine from our data the galaxy type we were 
detecting. The fact that two or three different acceptable models give
similar residuals is basically due to the rather low signal-to-noise
ratio at the wings of the quasars.

\subsection {Results. Apparent magnitudes of the quasar hosts galaxies}

Table 2 shows the final results of the fitting procedure for the 16
quasars with a detected extended component\footnote{Four of these
quasars were not considered as extended in Paper I, 
where we used a more restrictive
criterion for host galaxy detection, i.e. a galaxy contribution higher
than 30\% and $K_{\rm gal}<17.2$.  These sources are 0704+384,
0726+431, 0740+380C and 0849+424.}. Empty entries in the table
correspond to one poor fit and two fits in which the fractional
contribution of the galaxy was lower than 12\% (in these two cases
also $K_{\rm gal}>17.9$. The table lists the residuals (rms in mags)
and the parameters $K_{\rm gal}$ and $K_{\rm QSO}$ for these models,
along with the residuals for the gaussian model for comparison.  The
two last columns of Table 2 give the maximum dispersion in $K_{\rm
gal}$ and $K_{\rm QSO}$ between the different acceptable models for
each source (excluding the two-component models giving rms similar to
a pure gaussian model: King model for 0739+397, 0918+381 and 1111+408,
and $r^{1/4}$ model for 1315+396). The average dispersion between the
galaxy magnitudes derived from the acceptable models is 0.25 mag, the
median is also 0.25 mag and the larger discrepancy is 0.6 mag. For
$K_{\rm QSO}$ the average dispersion is 0.3 mag, the median is 0.2 mag
and the largest discrepancy is 1.1 mag, for 0756+406.

There is in general a good agreement
between the values of $K_{\rm QSO}$ and $K_{\rm gal}$ obtained for 
the different acceptable models. The similarity of $K_{\rm gal}$ for 
these models is a natural consequence of the way it is
measured, subtracting from the total light the QSO contribution
obtained from the model, which is roughly similar for the different
acceptable models. The average dispersion of $K_{\rm QSO}$ and 
$K_{\rm gal}$ for the different acceptable models, of typically
0.3 mag, is similar to the average rms of the acceptable models, which
is about 0.35 mag.

For the discussion on the next sections we have adopted as $K_{\rm
QSO}$ and $K_{\rm gal}$ the values for the model with the minimum rms.
The error in the apparent magnitude of each component has to be
obtained from the quoted rms of the fit and the dispersion
due to the selection of one particular model, since for all cases there are at
least two acceptable models, with roughly similar rms. Considering
these errors in quadrature, the average error for the apparent
magnitudes of the host galaxies is $\sigma (K_{\rm gal})=0.4 \pm 0.1$,
and for the QSOs, $\sigma (K_{\rm QSO})=0.4 \pm 0.2$ (excluding the
large error of 0756+406 from the average). The errors for these parameters 
correspond to the data dispersion.

We have obtained images of the galaxies subtracting the fitted
point-like sources (obtained from the minimum rms model) from the
deconvolved images.  These images were then convolved with the
original PSF so that they could be compared with the original quasar
images, and we call them the 'restored' images.  Figure 3 shows, for
each object, contour plots of the original quasar image and the
'restored' host-galaxy image, and their corresponding surface
brightness profiles, obtained as explained in Sect. 3.3 . The solid
line on the brightness profile of the quasar shows the PSF profile rescaled 
to the quasar peak flux.

For 8 of the extended objects some morphological distorsions or
peculiarities are apparent in both the original quasar images and the
restored galaxy images. These are radial elongations, tails,
distorsions, etc.  However, since the signal-to-noise of these features 
is rather low we
should take them with caution. Other sources (extended or
not) have nearby objects in the field, but these are faint and we
cannot confirm or reject their association with the radio source. All
these possible companion objects were carefully masked before the
surface photometry analysis was performed, and therefore they do not
contaminate the surface brightness profiles.

\section {K-z RELATION OF HOST GALAXIES}

The apparent $K$ magnitudes of the sixteen detected B3-VLA quasar hosts are
listed in Table 3, along with their errors, and the nominal ratio of
galaxy-to-total emission $L_{\rm gal}/L_{\rm tot}$.  The sixteen quasars
with detected hosts have redshifts ranging from $z=0.6$ to $z=2.3$, $L_{\rm
gal}/L_{\rm tot}$ ratios from 20 to 80\% and galaxy magnitudes $K_{\rm
gal}=15.2-17.8$, with the latter value roughly corresponding to the
detection limit.

In Figure 4 we plotted the apparent $K$
 magnitudes versus redshift of the detected B3-VLA quasar hosts and the
 corresponding lower limits for those classified as point sources.
For comparison we also plotted the $K$ magnitudes of
the RQ hosts detected by L92 and T96.
L92 galaxy magnitudes were corrected to account for the galaxy flux missed
in the inner-most 2 arcsec$^2$ area due to the PSF substraction.
Following the author's estimations of a few tenths of magnitude
correction, a correction of 0.4 magnitudes was applied. 
The brightest cluster galaxies (hereafter BCGs) in the redshift range 
$0.5<z<1.0$ studied by Arag\'on-Salamanca et al. (1993) are also shown.
Although the classification of a
galaxy as BCG is a relative one - the brightest of the cluster - these
galaxies are similar in most properties to gEs.

The thin continuous curve in Fig. 4 shows the $K-z$ relation for 3CR and
B2-1Jy class radio galaxies obtained by Lilly et al. (1985) and Lilly
(1989), and the superimposed vertical line marks its dispersion, of 0.4 mag
in the redshift range $1<z<2$. The dotted curve shows the $K-z$ relation
expected for a passively-evolving old and luminous gE, in which all star
formation has taken place during an initial burst, followed by no further
star formation. This model reproduces very well the $K-z$ relation found by
Lilly et al. (1985) and Lilly (1989) for RGs, and in fact this agreement
has been traditionaly one of the main arguments for the interpretation of
RGs as old passively evolving gEs.  The modelled $K-z$ relation was
obtained using Bruzual (1983) $c$-model and the recent implementation of
their code GISSEL (Galaxy Isochrone Synthesis Spectral Evolution Library,
Bruzual \& Charlot 1993, new version of 1995).  The $c$-model assumes a
constant star formation for an initial period $\tau$ and zero star
formation thereafter.  We used $\tau=1$ Gyr, $M_{\rm K}=-25.8$ - which is
the absolute magnitude of a BCG/gE for this cosmology (Thuan \& Puschell
1989; Arag\'on-Salamanca et al. 1993) - and $z_{\rm for}=10$, which for the
adopted cosmology corresponds to an age of 12.6 Gyr for a present day
galaxy. For this value of $z_{\rm for}$ the assumed star formation, lasting
$\tau=1$ Gyr, would have concluded at $z=3.5$. The initial mass function
used was a Salpeter IMF (1955), with $M$ from 0.1 to 125 $M_\odot$.

In a recent work Eales \& Rawlings (1996) reported a slightly fainter
$K-z$ relation for their sample of B2-1Jy and 6C RGs. This relation is
plotted as a dashed curve in Figure 4 and has a dispersion around 0.6
mag in the redshift range $1<z<2$. As noted by the authors, this
relation is well matched with a $c$-model with similar parameters
$\tau=1$ Gyr and $M_{\rm K}=-25.8$, but assuming no stellar
evolution. Since the stellar population must be evolving to some
extent, the match with a non-evolving model is explained by the
authors suggesting that the effect of stellar evolution is cancelled
out by some process which makes the galaxy luminosity increase with
age, as for instance the scenario of hierarchical clustering. The
dash-dotted curve shows the no-evolution model. The thick curve shows
the average $K-z$ relation for Lilly et al. and Eales \& Rawlings. The
former sample includes 3CR RGs with $S_{408}>5$ Jy and B2-1 Jy sources
with $1 {\rm Jy}<S_{408}<2$ Jy.  The RGs in the sample studied in
Eales \& Rawlings have flux densities in the range $0.8 {\rm Jy}< S_{408}<
1.6$ Jy. The flux densities of the B3-VLA quasars occupy an intermediate
position between the two samples, with a median flux density around 2
Jy. Obviously, a better comparison of the host luminosities with those 
of RGs will be possible when $K$-band imaging of a complete sample of 
RGs selected from the B3 catalogue  becomes available.

At low redshift ($z<1.0$) there is a good agreement between the
location of the detected B3 quasar hosts and RGs/gEs, as well as with
the BCGs studied by Arag\'on-Salamanca et al. (1993).  There is one
quasar at $z\simeq 0.5$, classified as point source, for which the
upper limit for the host luminosity is much fainter than the typical
luminosity of RGs and detected quasar hosts at these redshifts. A
possible explanation for the inferred low luminosity is that the
quasar is hosted by a compact galaxy, which could not be separated
from the nucleus in our analysis.  At high redshifts ($z>1.0$) the
detected B3-VLA quasar hosts tend to be brighter than RGs/gEs, and
this trend is also found, although to a lesser extent by L92.  We
note, however, that our detection limit, $K\simeq 17.9$, prevents the
identification of galaxy hosts near the expected location for RGs at
$z~\mayeq~1.0$, and therefore only the galaxies brighter than this
relation are expected to be detectable in our survey at these
redshifts. In fact, eight of the sources with $z>1.0$ in Fig. 4,
comprising two weak galaxy detections and six lower limits, have
galaxy magnitudes (or limits) consistent with the expected values for
RGs. But the eight detected host galaxies with $z>1.0$ (the two weak 
detections excluded) are clearly more luminous than typical RGs, 
deviating in seven cases more than 2$\sigma$ relative
to the $K -z$ relation by Lilly et al. (1985). The latter detections
indicate that the B3 quasar hosts can reach higher luminosities than
typical RGs/gEs, although the data imply a similar fraction of hosts
having $K$ magnitudes consistent with those of RGs/gEs.

A possible explanation for the largest luminosity of the detected
hosts at $z>1$ is that they are gEs, similar to RGs, except for a
larger mass, i.e.  intrinsic luminosity.  This interpretation would
imply that RQ hosts have a range of possible masses, since as shown in
Fig. 4, the low redshift RQ hosts and some high redshift lower limits
are consistent with having a mass similar to typical RGs/gEs. As we
shall see below, under the assumption of passive stellar evolution,
i.e. stellar evolution with no additional star formation after an
initial burst, the largest luminosities obtained are not unreasonable,
since they correspond to the bright end of the luminosity functions
derived from galaxy surveys selected in the $K$ band.

A second possibility is that these galaxies are gEs similar to RGs, but
formed more recently. The long-dash-short-dash curves on Fig. 4 show the
$K-z$ relation for $c$-models with $\tau=1$ Gyr, $z_{\rm for}=2,3,5$ and
assumed stellar evolution.  Especially the model with $z_{\rm for}=3$ shows
a good agreement with the data.  For these models, the star formation,
lasting 1 Gyr, ends at redshifts $z=1.2,2$ and $2.8$, which roughly
correspond to the minima in the $K-z$ curves. The large $K$ band luminosity
in these models is due to the contribution of massive red giants and
supergiants. Since RQs do exist at larger redshifts, in these models
different RQs would require different formation epochs. A problem with
these models is that they appear to be inconsistent with the total $B$
magnitudes measured for some of the quasars. For instance, using the model
with $z_{\rm for}=3$, the two well defined galaxies (filled circles) with
$z>2$, 0756+406 and 1142+392, would have blue colours, $B-K \simeq 2$ mag,
implying $B_{\rm gal}$ about 19.1 and 18.7 respectively, and these values
are lower in the first case and similar in the second case to the total
quasar magnitudes in $B$ (including the central QSO), of 19.9 and 18.8
magnitudes respectively.

A third possibility is that they are old gEs, similar in mass and age
to RGs, but undergoing a large late starburst, probably related to the
nuclear activity. The excess of $K$ emission of the luminous galaxies
relative to the $K-z$ relation for typical RGs is roughly $\Delta K \simeq 1.5
$ mag and the derived starburst contribution would be $K_{\rm SB}
\simeq 17$ mag. Assuming a flat spectrum during the starburst (Bruzual
\& Charlot 1993; Lilly 1989), corresponding to $B-K \simeq 2$ mag, 
the expected contribution in the $B$-band would be typically 
$B_{\rm SB} \simeq 19$ mag. Six of the eight RQs harboured by
luminous galaxies have total $B$ magnitudes (including the nuclear
contribution) larger (four cases) or similar (two cases) than the
expected value for $B_{\rm SB}$, rejecting this model as a possible
general interpretation for the luminous galaxies.

We have calculated the absolute $K$ magnitudes at $z=0$ of the B3-VLA
quasar host galaxies, assuming that through the whole redshift range
the $K$ emission is produced by a mature population of
passively-evolving stars. Although models involving young galaxies
could be valid for some sources, they can be disregarded as a general
interpretation, since they predict for the youngest sources a galaxy
flux in $B$ which is larger than the total flux measured for the
quasar (nucleus included). The same occurs for the model in which {\it
all} the $K$ excess relative to a typical gE is attributed to young stars,
since it also would predict $B$ fluxes for the galaxies which would be
generally larger than observed.  An additional problem of these models
is that they include unknown parameters such as the epoch and duration
of the star formation period relative to the age of the galaxy, which
would strongly affect the derived absolute magnitudes. In any case,
for the galaxies with some degree of current star formation or formed
at $z<10$ the derived absolute magnitudes at $z=0$ under the
hypothesis of passive stellar evolution would be lower-limits.  
The absolute
$K$ magnitudes were calculated using the $k$ and evolutionary
corrections obtained from GISSEL for the $c$-model with $\tau = 1$ Gyr
and $z_{\rm for}=10$.  $k$-corrections, evolutionary corrections and
absolute $K$ magnitudes of the sixteen galaxies are listed in Table
3. For the redshifts of the hosts, from $z=0.6$ to $z=2.3$, the $k+e$
corrections range from about --1.10 to about --2.5 mag.

In Figure 5a we plot the absolute magnitudes so obtained, $M_{\rm K}$
versus redshift for the detected RQ hosts in our work and those in T96 and
L92. The curve corresponds to our maximum limiting magnitude for the
detection of a host galaxy, $K=17.9$.  The absolute magnitudes of the B3
quasar hosts with $0.6 < z < 1.0$ are in the range $-26.0 ~\meneq ~M_{\rm
K} ~\meneq ~-24.7$, showing a good agreement with the average absolute
magnitudes of the RQ hosts in T96 sample between apertures $r=12$ arcsec
and $r=\infty$, and with those of the $0.5<z<1.0$ BCGs of
Arag\'on-Salamanca et al. (1993), with $M_{\rm K}=-25.7 \pm 0.3$, as
expected from Fig. 4. For comparison Thuan \& Puschell (1989) found a
slightly lower value $M_{\rm K}=-26.0$ for BCGs at $z<0.1$.  We note that
the faint end of the absolute magnitude distribution for our data, $M_{\rm
K} \simeq -24.7$, may be related to the detection limit in our analysis.

For this model, the detected RQ hosts in our sample with $z>1$ have absolute
magnitudes in the range $-27.5<M_K<-26.5$. The corresponding absolute
magnitudes for L92 RQ hosts would be slightly fainter, in the range $-27.0 <
M_K <-26.0$ (excluding the galaxy with a large error), 
but their faintest sources would be undetectable or only
marginally detectable in our work. The bright absolute magnitudes
measured for the high redshift sources in our work $-27.5<M_K<-26.5$
correspond to the bright end of the $K$-band luminosity function obtained
for the E/S0s galaxies in the Anglo Australian Redshift Survey
(Mobasher et al. 1986, redshifts in the range $0<z<0.11$), and for the recent 
$K$-band surveys by Glazebrook et al. (1996), up to $z\sim 0.8$ and Gardner
et al. (1997) up to $z\sim 0.3$.

Considering the whole redshift range, the absolute $K$ magnitudes of
RQ hosts show some correlation with redshift, in the sense that lower
redshift hosts ($z<1$) do not reach as large luminosities at $K$ as
can be reached by high redshift hosts. This result has the implicit
assumption that the $K$ emission arises from a mature elliptical 
galaxy, which appears to be the most favoured model for the $B$
and $K$ band data. Although star formation cannot account for {\it all} the 
excess of $K$ emission of the $z>1$ galaxies, relative to typical gEs,
it may produce part of this excess.

In Figure 5b we plot $M_{\rm K}$ versus redshift for the same sources,
but using the no-evolution model (only $k$-correction), which matched
the sample of RGs studied by Eales \& Rawlings (1996). For the low
redshift sources ($0.6<z<1.0$) the absolute magnitudes of the B3
quasar hosts are in the range $-26.5 ~\meneq ~M_{\rm K} ~\meneq
~-25.2$, showing a better agreement with the value $M_{\rm K}=-26.0$
found by Thuan \& Puschell (1989) for BCGs at $z<0.1$. However, the
absolute magnitudes derived for the the high redshift sources for this
model, in the range $-29.0 ~\meneq ~M_{\rm K} ~\meneq ~-27.5$, are by
far more luminous than the brightest galaxies known from $K$-band
surveys.  For this reason we believe it is more
appropriate to use the model with stellar evolution, 
which yields absolute magnitudes for the RQ hosts
similar to those of the brightest galaxies known.

So far we have tried to explain the high $K$-band luminosity of some of the
RQ hosts galaxies using models of an ideal unperturbed galaxy for
which we varied the mass or luminosity, the formation epoch or
introduced renewed star formation. An alternative explanation to these
high absolute magnitudes comes from the suggestion that the activity
in radio sources is triggered by merging processes involving massive
galaxies (see the reviews by Barnes \& Hernquist, 1992, and others in
Shlosman, 1994).  There is in fact circumstancial evidence, up to $z
\simeq 0.5$, that a large fraction of RQ hosts undergo tidal
interactions/merging processes (Smith et al. 1986; Hutchings 1987;
Disney et al. 1995; Hutchings
\& Neff 1997). In this case, the very process of galaxy interaction would
naturally allow for a wide range of luminosities, depending on the
luminosities of the interacting/merging galaxies. Some 
induced star formation could be also related to the 
interacting/merging process. 

It is crucial to obtain high spatial resolution images of the detected
B3-VLA quasar hosts to study their morphologies in more detail.  Such
images would allow to better determine the contribution of extended
emission due to synchrotron and/or alignment effect, which is estimated to
be low in the infrared for $z\sim 1$ RGs and RQ hosts.  The study of the 
optical-infrared SED of the host galaxies is also necessary to better
constraint the evolutionary models.

\section {$B-K$ COLOURS OF THE QUASARS}

In Paper I the study of the integrated $B-K$ colours of this sample of
B3-VLA quasars was presented. The colours were obtained using the $B$
magnitudes of the quasars taken from the catalogue of objects on the POSS-I
blue plates generated by the APM (accuracy 0.3 mag rms) and the $K$
magnitudes measured from the images (both in Table 1).  The distribution of
$B-K$ colours for the B3 quasars (shown in Fig. 1a of Paper I) is similar
in breadth to that found by Webster et al. (1995, hereafter W95) for
flat-spectrum ($\alpha_{2700}^{5000} > -0.5$) radio-selected quasars (their
Fig. 1b), except for the lack of extreme red colours $B-K > 6$. W95
interpreted the dispersion in the $B-K$ colour of their quasars ($1<B-K<8$)
in terms of dust reddening, implying an extinction in the blue of several
magnitudes for a substantial fraction of the quasars.

In Paper I we provide evidence that for most of the 
red B3-VLA quasars, the red colour is due to additional light in $K$ (starlight
or synchrotron emission) rather than a deficit in $B$ due to dust
extinction. The arguments for this interpretation are explained below
and they are illustrated in figures 1a ($B-K$ histogram) and 2 ($K-z$
diagram) of Paper I.  Many of the reddest B3 quasars have non-stellar
images in $K$, consistent with the presence of underlying galaxies,
and indeed many of the B3 quasars have $K$ magnitudes which are not
much brighter than their host galaxies are expected to be, assuming
giant ellipticals with $M_K \sim -26$. 
Although the $B-K$ distribution for the flat and steep-spectrum quasars in
our sample is not significantly different, of the six sources with the
flattest spectra ($\alpha > -0.3$) only one has $B-K < 3.5$. 
These red colours are consistent with the presence of
an enhanced synchrotron emission due to relativistic beaming, peculiar
to flat-spectrum quasars, which extends to the optical and infrared
(Bregman et al. 1981; Rieke et al. 1982; Browne \& Murphy 1987).
Serjeant \& Rawlings (1996) have recently argued that such non-thermal
emission may explain the red colours found by W95. 
The lack of quasars with $B-K>6$ (W95 found several) in the 
B3-VLA Quasar Sample, dominated by steep-spectrum sources, is consistent with
this interpretation. Additional
support to the hypothesis that the reddening was due to a $K$ excess
came from the fact that two objects which stand out as being
particularly luminous in $K$ were also very red (see in Fig. 2 of
Paper I the objects with $B-K=5.6$ (flat) and $B-K=4.4$, both with
$K<14$).  These sources are shown in Fig. 1b of the present work
with underlined symbols.

Figure 6a shows the $B-K$ colour versus redshift for the 52 B3-VLA
quasars.  The distribution is similar to that found by W95, except for
the lack of very red quasars ($B-K > 6$). 
The average $B-K$ colour for the whole sample,
consisting predominantly of steep-spectrum quasars, is $B-K=3.3 \pm
1.1$.

In total we have found eighteen B3-VLA quasars for which the $K$
images show non-stellar profiles (see Sects. 3.2 and 3.3). For sixteen
of them the light distribution is well fitted by models consisting of
a QSO nucleus plus a host galaxy and the magnitudes of both components
could be determined (rms around 0.4 mag).  Twelve additional quasars
were classified as point sources from the analysis in Sect. 3.2, and
thus for these sources $K_{\rm QSO}=K$. Hence for these 28 sources we
can study the intrinsic $B-K$ colours of the QSO, provided that the
assumption can be made that most of the $B$ emission arises from the
QSO.  The model of an old elliptical galaxy ($z_{\rm for}=10$), which
is the model that better explains the properties of RQ hosts and RGs
(V\'eron-Cetty \& Woltjer 1990; Disney et al. 1995; T96; Ridgway \&
Stockton 1997; Lilly et al. 1985; Lilly 1989; Rigler et al. 1992; Best
et al. 1997), has $B-K>6.0$ for the whole redshift range of our data,
implying $B \sim 22$ for the brighter galaxies, with $K \simeq 16$,
and even fainter $B$ magnitudes for the rest. These $B$ magnitudes are
well below the typical total $B$ magnitudes of the B3-VLA quasars (see
Fig. 1), indicating that the relative galaxy contribution in $B$ would
be very small. Although the measured $B$ and $K$ fluxes rule out the
model of an old galaxy with a large late starburst and the model of a
young galaxy ($z_{\rm for}<5$), the data cannot reject, however, the
presence of some low-level star formation, producing some contribution
in $B$, similar to that found in FRII RGs (Heckman et al. 1986).  In
addition, some contribution in $B$ could arise from scattered QSO
light, a process which has also been observed in some RGs (e.g. di
Serego Alighieri et al. 1989).  In this section we study the intrinsic $B-K$
colours of the twenty-eight quasars with determined $K_{\rm QSO}$,
assuming that most of the quasar emission in $B$ arises from the
central QSO. The fact that the sources were identified in the POSS
plates as pointlike lends additional support to this assumption.
Hereafter $B-K$ will refer to the quasar colour and $B-K_{\rm QSO}$ to
the QSO colour. Figures 6b and 6c show respectively the $B-K$ colour
and $B-K_{\rm QSO}$ colour versus redshift for the twenty-eight
sources with determined $K_{\rm QSO}$.

For the redshift range of this quasar sample, $0.4<z<2.3$ the $K$ band
corresponds to rest-frame regions from 1.6 $\mu$m to 6700 \AA~ (or
$14.30 < {\rm log} ~\nu < 14.65$) and $B$ corresponds to 3000--1300
\AA~ (or $15.00 < {\rm log} ~\nu < 15.35$). The average SED of QSOs is
well fitted by a power law, with $S_\nu \propto \nu^\alpha$, with a
break at rest-wavelength around 1 $\mu$m (${\rm log} ~\nu = 14.5$), in
which the spectral index steepens from high to low frequencies
(Neugebauer et al. 1987; Sanders et al. 1989; Elvis et al. 1994).
Although this general shape is well established, the dispersion around
the average is rather large. The mean $\alpha_{\rm opt}$, at ${\rm
log} ~\nu > 14.5$, is estimated to be around $-0.2$ (Neugebauer et
al. 1987). For $\alpha_{\rm NIR}$, at ${\rm log} ~\nu < 14.5$,
Neugebauer et al. quote $-1.4$. The typical SED of a radio quasar
shown by Elvis et al. (1994, their Fig. 1) has a lower $\alpha_{\rm
NIR}$, closer to $-1$, which is also the value used by Sanders et
al. (1989).

Figure 7 shows the average rest-frame SED of QSOs using $\alpha_{\rm
NIR}=-1$ and $\alpha_{\rm opt}=-0.2$. The rest-frame wavelengths at
$B$ and $K$ for several redshifts are marked in the plot, using
circles for $K$ and squares for $B$. The redshifts are labelled.

For the quasars with $z>1.3$, both $B$ and $K$ sample the optical
range, i.e. ${\rm log} ~\nu > 14.5$. Assuming a constant spectral
index, $\alpha_{\rm opt}$, we would expect a constant $B-K_{\rm QSO}$
colour, which does not change with redshift.  The average $B-K_{\rm QSO}$
colour for the 11 QSOs in Fig. 6c with $z>1.3$ is $B-K=2.5\pm1.2$, which corresponds to
$\alpha_{\rm opt}=-0.22$, in good agreement with Neugebauer et
al. (1987) and Elvis et al. (1994). A dotted line indicates this
average colour in Fig. 6c.

As we move to $z<1.3$ in Fig. 7 the $K$ band starts to sample infrared
frequencies, below the frequency break at ${\rm log} ~\nu=14.5$, and
the slope corresponding to the $B-K_{\rm QSO}$ colour steepens,
i.e. the $B-K_{\rm QSO}$ colour turns redder (see the dashed lines in
Fig. 7).  This result is consistent with our data, since the $B-K_{\rm
QSO}$ colour for the QSOs with $z<1.3$ is in average redder, with
$B-K_{\rm QSO}=3.2\pm1.2$. A dotted line in Fig. 6c shows this average.  This
$B-K_{\rm QSO}$ colour corresponds to an spectral index
$\alpha=-0.64$, which is an upper limit to $\alpha_{\rm NIR}$, since
the $B$ band samples frequencies higher than the break.  (It is
obvious from Fig. 7 that $\alpha_{\rm NIR}$ is steeper than the slopes
shown as dashed lines).  Neugebauer et al. (1987) and Elvis et
al. (1994) give $\alpha_{\rm NIR}$ in the range $-1.4$ to $-1$,
consistent with our upper limit at $\alpha=-0.64$.

The effect of the removal of the host galaxy on the $B-K$ colours can be
appreciated comparing these colours in Figures 6b and 6c. 
Considering the extended sources only, the average $B-K$ colour is reduced
from 3.5 to 2.8 for $z<1.3$ and from 2.9 to 1.9 for $z>1.3$.

The $2\sigma$ dispersion of the $B-K_{\rm QSO}$ colours in Fig. 6c 
is of 2.4 both for the QSOs with
$z<1.3$ and $z>1.3$. This dispersion in the $B-K_{\rm QSO}$ colours
limits the amount of reddening to rest-frame dust extinctions $A_{\rm V} <
1.5$, $A_{\rm V} < 1$ and $A_{\rm V} < 0.8$, at $z=0.5$, $z=1.0$ and $z=2.0$
respectively, where we used the reddening law of Kinney et al. (1994) 
for the optical/UV and Rieke \& Lebofski (1985) for the infrared. These
$A_{\rm V}$ values should be taken as conservative limits, since one would
expect some of the spread in the $B-K_{\rm QSO}$ colours to be
intrinsic. The inferred extinctions are 
in agreement with the
values obtained for other quasar samples selected in radio, X-rays or
in the optical (Schmidt 1968; Smith \& Spinrad 1980; Netzer et
al. 1995; Boyle \& di Matteo 1995; Rowan-Robinson 1995, Baker 1997).  
Figure 6d shows the $B-K$ colours versus redshift for the quasars for
which we could not determine $K_{\rm QSO}$. The colour distribution is
rather flat and excluding the two
quasars showing spatial structure (plotted as squares, one of them
very red) the average $B-K$ colour of these sources is $B-K=3.0
\pm1.3$. For most of these sources the analysis of extension described
in Sect. 3 was not performed, since either the sources were observed
the second night or although observed the first night they had or
could have problems of guiding/focusing. Therefore we expect that some
of these sources also present some galaxy contamination at
$K$. 

\section {ABSOLUTE $K$ MAGNITUDES OF THE QUASARS AND CORRELATIONS WITH RADIO
LUMINOSITY}

We have calculated the $K$ absolute magnitudes of the nuclear components
for the twenty eight quasars for which $K_{\rm QSO}$ could be determined. In
order to correct to rest-frame $K$ absolute magnitudes we assumed a
spectral index $\alpha=-1$ in the near infrared, although for $z>1.3$ the
value could be slightly higher (more flat, see Sect. 5 and Fig. 7). The
assumption of a spectral index $\alpha=-1$ implies a null $k$-correction.
The $K$ absolute magnitudes of the QSOs are listed in Table 4. The last
column of the table indicates whether the quasar was pointlike or had
resolved extended emission. In Figure 8 we show the diagram $M_{\rm K,QSO}$
versus redshift for these quasars.  The data clearly show luminosity
evolution with redshift in the $K$ band. The average $M_{\rm K,QSO}$ for
the sources at $z<1$ is $M_{\rm K,QSO}=-27.6 \pm 0.9$, but values as high
as $M_{\rm K,QSO}=-30$ are reached at $1<z<2$.  It is interesting to note
the location on the diagram of the quasar 1144+402, with $z=1$ and $M_{\rm
K,QSO}=-30.8$. This quasar is pointlike, has a flat radio spectrum with
$\alpha=0.06$, a very bright nucleus at $K$, and a very red colour
$B-K=5.6$. This is the flat-spectrum quasar appearing underlined in
Fig. 1b. It is evident that the red colour of this source is due to a $K$
excess rather than a defect at $B$, and the $K$ excess is probably related
to the flat radio spectrum. The flat-spectrum sources appear to have
brighter nuclei at $K$ in this diagram than the steep-spectrum ones.

In any radio selected flux-limited sample there is a strong ''artificial''
correlation between radio luminosity and redshift. Hence, it is possible
that the brighter nuclear luminosities at $K$ at redshifts $z>1$ are
related to the stronger radio power at these redshifts. Figure 9 shows the
rest-frame luminosity at 1.4 GHz versus redshift for the B3-VLA quasars
studied in this work. We used the observed total flux densities at 1.4 GHz
and assumed a power law with the spectral index $\alpha_{408}^{1460}$ given
in Table 1 to make the $k$-corrections, of the form ${\rm log}
~(1+z)^{-1-\alpha}$.  Figure 9 clearly shows the artificial correlation
between radio power and redshift mentioned above.

In Figure 10 we plotted the radio power $P_{1.4~\rm GHz}$ versus $M_{\rm
K,QSO}$ for the B3-VLA RQs with measured $K_{\rm QSO}$. We found in fact a
correlation between the two parameters, with a Spearman coefficient of
$-0.57$ (significance level of 99.85\%). This correlation improves
considering only the steep-spectrum sources, as some of the flat-spectrum
sources deviate towards higher nuclear luminosities at $K$. The Spearman
coefficient of the correlation for steep-spectrum sources is $-0.71$ with a
significance level of 99.97\% and corresponding to $L_{\rm radio} \propto
L_{\rm K,QSO}^{0.75\pm 0.15}$. This correlation remains significant in the
flux-flux plane (Fig. 11a), and the latter is not induced by distance
effects, as low and high redshift RQs are equally distributed, each group
showing the same correlation and spread as the whole sample (Fig. 11b). The
flux-flux correlation for steep-spectrum sources has a Spearman coefficient
of $0.72$ with $99.96$\% significance.  Therefore the data favour a direct
link between the $K$ flux from the nucleus and the synchrotron radio
emission, which is more tight for the steep-spectrum quasars. The
correlation $M_{\rm K,QSO} - z$ shown in Fig. 8 is weaker than the
correlations $P_{1.4~\rm GHz} - M_{\rm K,QSO}$ and $P_{1.4~\rm GHz} - z$
and it is most likely induced by these two correlations.

Figure 12 shows a plot of $P_{1.4~\rm GHz}$ versus $M_{\rm K,gal}$ for the
sixteen B3-VLA quasars for which the galaxy component was measured, via
light profile fitting.  $M_{\rm K,gal}$ was calculated under the assumption
that the whole $K$ emission was produced by the old stars of a passively
evolving galaxy.  A Spearman test for $L_{\rm radio} - L_{\rm K,gal}$ gives
only a weak correlation ($r=-0.49$, 94\% confidence level), corresponding
to $L_{\rm radio} \propto L_{\rm K,gal}^{0.6\pm 0.3}$, and the correlation 
dissapears in the flux-flux plane (Fig. 13, $r=0.39$, 87\%
confidence level). We believe that the weak 
$P_{1.4~\rm GHz}$-$M_{\rm K,gal}$ correlation is real; although the magnitude limit for the detection
of the host galaxies could account in part for the lack of sources in the
upper-left region of the $P_{1.4~\rm GHz}$ versus $M_{\rm K,gal}$ diagram, the
absence of bright galaxies at low radio powers clearly indicates a real
$L_{\rm radio}- L_{\rm K,gal}$ trend, in the sense that the brighter
galaxies host the more powerful radio quasars.  As we did for the $M_{\rm
QSO}-z$ relation, we interpret the trend $M_{\rm gal}-z$ in Fig. 5a as
induced by the relations $M_{\rm gal}-P_{1.4~\rm GHz}$ and $P_{1.4~\rm
GHz}-z$. The parameters $M_{\rm K,QSO}$ and $M_{\rm K,gal}$ appear to be
marginally correlated ($r=-0.45$, 92\% confidence level) although the
correlation dissapears in the $K_{\rm QSO}-K_{\rm gal}$ plane (Figures 14
and 15).

The trend $P_{1.4~\rm GHz}$-$M_{\rm K,gal}$ indicates that powerful
radio quasars inhabit the most massive galaxies. In particular, all the
bright RQ hosts, with $M_{\rm K,gal}<-26$, have radio powers $P_{1.4~\rm
GHz} > 10^{27.5}$ WHz$^{-1}$. A similar tendency was found by Yates,
Miller \& Peacock (1986) for RGs. These authors selected a sample of FR-II
3C RGs and concluded that for the more powerful sources ($P_{\rm 178 MHz} >
10^{27}$ W Hz$^{-1}$) there is a correlation between the $K$ absolute
magnitude of the host galaxy and radio power.  They suggested that the
correlation is caused by the increase of stellar luminosities by
cannibalism in the most powerful radio galaxies.  As suggested by Baum,
Heckman \& van Breugel (1992) and Baum, Zirbel \& O'Dea (1995) mergers at
earlier epochs would have provided high angular momentum gas to fuel
accretion onto the nucleus, producing an FR-II radio source or a quasar. If
we translate our data to Figure 1 in Yates et al. (1986) we observe that
the quasar hosts lie in the same region that the most powerful RGs and in
fact both would be indistinguishable. This gives support to unification
schemes of radio sources, where FR-II RGs are the parent population of
radio quasars and both are expected to have similar relationships between
radio power and host galaxy properties.

\section{SUMMARY AND CONCLUSIONS}

We have obtained $K$-band images of a representatitve sample of 52
quasars with $0.3<z<2. 8$, $S_{\rm 408 ~MHz}>0.5$ Jy and radio powers
$P_{\rm 1.4 ~GHz} \simeq 10^{27}-10^{28}$ WHz$^{-1}$, whose analysis yield the
following conclusions.

For sixteen out of 28 sources for which the images allowed a detailed
morphological analysis, resolved extended emission was detected around
the central QSO, with estimated apparent magnitudes in the range $K
\simeq 15.5-17.5$. The redshifts of these QSOs with extended emission
are in the range from 0.6 to 2.3. Interpreting this ``fuzz'' as
starlight emission from the host galaxy, its location in the $K-z$
diagram for the low-redshift quasars, with $z ~\meneq ~1$, is
consistent with these quasars being hosted by galaxies similar to
gEs/RGs at these redshifts.  At high redshifts, $z ~\mayeq ~1$, the
extended emission reaches higher luminosities than those 
found for typical gEs/RGs at these
redshifts, a trend which was also found although to a lesser extent
by L92.  Since at $z ~\mayeq ~1$ a typical gE has apparent magnitudes in
$K$ near the detection limit for our analysis, only the most
luminous galaxies are detectable in our work at these
redshifts.

We have considered three possible explanations for the large
luminosities of the detected host galaxies with $z ~\mayeq ~1$: mature
elliptical galaxies with masses larger than a typical gE; young
elliptical galaxies, formed at $z \simeq 2-4$, with present-day $K$
luminosity similar to gEs; and mature galaxies similar in mass to gEs
but undergoing a recent starburst. The two last models cannot be taken
as a general interpretation of the luminous galaxies, since they
predict for many of the sources a larger $B$ flux from the galaxy than
the total $B$ flux measured for the quasar, including host and
nucleus. The most appropriate model to account for both the $K$ and
$B$ band data for the majority of the sources is that of a mature
passively evolving elliptical galaxy with a range of possible
masses. Using this model we derive a present-day absolute $K$
magnitude, $M_{\rm K} \simeq -26$, for the B3-VLA quasar hosts at $z
~\meneq ~1$, similar to the typical values found for RGs. The absolute
magnitudes we obtain for the high redshift B3-VLA quasar hosts,
$M_{\rm K} \simeq -27$, correspond to the bright end of the $K$-band
galaxy luminosity functions obtained from galaxy surveys selected in
this band (Mobasher et al. 1996; Glazebrook et al. 1996; Gardner et
al. 1997). Rephrasing L92, `` the host galaxies of RQs at any redshift
are probably a subset of the most massive galaxies in existence''.  A
possible explanation of these large luminosities is that the host
galaxies of RQs are the product of a past merger event.  This process
would also contribute to the triggering of the nuclear activity.

From a quantitative analysis of the $B-K_{\rm QSO}$ colour of the
B3-VLA quasars for which the nuclear component could be measured, we
confirmed our previous conclusion, based on the integrated quasar
$B-K$ colours (Paper I), that the $B-K$ colours of the
nuclear component show a small dispersion, of about 2.4 mag. This
small dispersion in the $B-K$ colour imply a low reddening, $A_{\rm V}
< 1.0$ at $z \simeq 1$ similar to that found for other quasar samples
selected in the optical, radio, or X-rays.

We have found a correlation between radio power and nuclear $K$-band
luminosity, indicating a direct link between the nuclear infrared
emission and the radio synchrotron emission. The correlation is more
tight for the steep-spectrum sources (99.97\% significance). In
addition a relation between radio power and infrared luminosity of the
host galaxy is found, in the sense that the most luminous (massive)
host galaxies contain the most powerful RQs.  It would be interesting
to increase the number of B3-VLA quasars with measured hosts, through
obtaining new $K-$band images, to confirm this trend. A similar result
was found for powerful FR-II RGs by Yates et al. (1986), who suggested
that the trend is caused by the increase of stellar luminosities by
cannibalism in the most powerful RGs. The similarity of the relations
between radio power and galaxy luminosity found for FR-II RGs and RQs
supports the unification between the two populations.

\vspace{1cm}

\noindent {\em Acknowledgements. }

We have made use of the APM Catalogues. 
We thank the referee, R. Antonucci, for valuable comments and
suggestions which improved the paper. We also thank 
S. Charlot for providing us with the
GISSEL package, and L. Cay\'on and I. Ferreras for their help in obtaining the 
$k$ and evolutionary corrections from the GISSEL models.
The William Herschel Telescope is operated by the Royal Greenwich
Observatory at the Spanish Observatorio del Roque de Los Muchachos of
the Instituto de Astrof\'\i sica de Canarias on behalf of the Science
and Engineering Research Council of the United Kingdom and the
Netherlands Organization for Scientific Research.  
Financial
support was provided under DGICYT project PB92-0501, DGES project PB95-0122
and by the Comisi\'on
Mixta Caja Cantabria-Universidad de Cantabria.

\newpage

\figcaption{Distribution of $B$ and $K$ magnitudes versus redshift for the 52
B3-VLA quasars studied in this work. Crosses correspond to flat-spectrum
sources ($\alpha_{408}^{1460} > -0.5$, $S_\nu \propto \nu^\alpha$).
\label{fig1}}

\figcaption{Surface brightness profile of B3 1111+408 obtained from the
deconvolved image. The solid line is the best fit with a two-component
model consisting of a nucleus (dash-dotted line) and an elliptical galaxy
(dashed line).
\label{fig2}}

\figcaption{Contour maps and surface brightness profiles of the extended
sources. For each quasar panel (a) is a contour map of the original image;
panel (a.1) is its corresponding brightness profile (dots) plotted with the
PSF (solid line); (b) is a contour map of the nucleus-subtracted image 
and (b.1) its corresponding brightness profile. Orientation in the contour
plots is north to the top and east to the left and the 
distance between tick marks is 2.7 arcsecs.
Intensities are given in counts and the spacial axes are in pixels.
\label{fig3}}

\figcaption{$K-z$ diagram for the B3-VLA quasar hosts, and other galaxies
or RQ host galaxies from the literature.  Circles show the detected B3-VLA
hosts (empty circles corresponding to detections with weaker significance)
and filled triangles lower limits to the host magnitudes.
L92 RQ hosts are indicated as pentagons and squares represent
T96 best fitted RQ hosts. Empty squares correspond to 12 arcsec aperture
magnitudes and filled squares to total magnitudes estimated by the authors
integrating the best-fit models to $r=\infty$. The BCGs in
Arag\'on-Salamanca et al. (1993) are shown as stars.
The thin continuous curve is the $K-z$ relation of Lilly et al., the short
dashed curve is the same relation for B2-1Jy/6C RGs of Eales \& Rawlings,
and the thick curve is the average of both relations. The remaining curves
correspond to models. The dotted curve is Bruzual \& Charlot (1993) {\it
c}-model for $z_{\rm for} = 10, \tau = 1$ Gyr and $M_{\rm K} = -25.8$, the
dashed-dotted curve is a similar model but considering no stellar
evolution, and the long-dash-short-dash curve shows models similar to the first
one but using $z_{\rm for} = 2, 3, 5$.
\label{fig4}}

\figcaption{Present day absolute $K$ magnitude, $M_{\rm K,gal}$, versus
redshift for the detected and measured B3-VLA quasar host galaxies and
other galaxies or RQ host galaxies from the literature, assuming passive
stellar evolution (a) or no-evolution (b). Symbols are the same as for
Fig. 4.  The curve corresponds to our maximum limiting magnitude for the
detection of a host galaxy, $K=17.9$.
\label{fig5}}

\figcaption{$B-K$ colours versus redshift for the B3-VLA quasars.
Flat-spectrum sources are indicated with crosses. Triangles show the
pointlike sources and empty squares the two quasars showing spatial
structure on the images. The quasars with a fitted galaxy component are
shown in panels (b) and (c) as filled circles (significant detection) or
empty circles (weak detection).
The dotted lines in panel (c) show the
average $B-K_{\rm QSO}$ colours for redshifts $z<1.3$ and $z>1.3$.
\label{fig6}}

\figcaption{Standard mean SED of a QSO using 
$S_\nu \propto \nu^\alpha$ with $\alpha_{\rm opt}=-0.2$,
$\alpha_{\rm NIR}=-1$ and the break at ${\rm log} ~\nu=14.5$ (
Neugebauer et al. 1987, Sanders et al. 1989, Elvis et al. 1994).  
Filled circles show the rest-frame frequencies observed in
the $K$ band at redshifts $z=0$, $z=0.5$, $z=1.5$ and $z=2.0$.  Empty
squares show the same data for the $B$ band.
\label{fig7}}

\figcaption{$M_{\rm K,QSO}$ versus redshift for the 28 quasars with
determined $K_{\rm QSO}$. Symbols are similar as for figs. 4-6.
\label{fig8}}

\figcaption{Radio power, $P_{\rm 1.4 ~GHz}$, versus redshift for the 52
B3-VLA quasars in this work. The curve shows the radio power $P_{\rm 1.4
~GHz}$ corresponding to a flux limit $S_{408}=0.6$ Jy and the average
spectral index of the sample $\alpha_{408}^{1460}=-0.79$.
\label{fig9}}

\figcaption{$P_{\rm 1.4 ~GHz}$ versus $M_{\rm K,QSO}$ 
for the 28 quasars with determined $K_{\rm QSO}$.
\label{fig10}}

\figcaption{Total measured radio flux at 1.4 GHz versus $K_{\rm QSO}$ 
for the 28 quasars with determined $K_{\rm QSO}$. For panel (a) the symbols 
are similar to those used for figs. 4-6. For panel (b) two 
different symbols are used to separate the sources in two redshift bins.
\label{fig11}}

\figcaption{$P_{\rm 1.4 ~GHz}$ versus $M_{\rm K,gal}$ 
for the sixteen quasars with determined $K_{\rm gal}$.
\label{fig12}}

\figcaption{
Total measured radio flux at 1.4 GHz versus $K_{\rm gal}$ 
for the 16 quasars with determined $K_{\rm gal}$. For panel (a) the symbols 
are similar to those used for figs. 4-6. For panel (b) two
different symbols are used to separate two redshift bins.
\label{fig13}}

\end{document}